\documentclass{ws-procs9x6}

\font\mymsbm=msbm10 at 10pt

\def\myb{\boldsymbol}
\def\half{{\textstyle \frac 1 2}}
\def\br{{\myb r}}
\def\bR{{\myb R}}
\def\bd{{\myb d}}

\def\bk{{\myb k}}

\renewcommand{\thetable}{\Roman{table}}

\def\bD{{\myb D}}
\def\bw{{\myb w}}
\def\bW{{\myb W}}
\def\bP{{\myb P}}
\def\bQ{{\myb Q}}
\def\bn{{\myb n}}

\def\T{{\rm T}}
\def\cal#1{\mathcal{#1}}
\begin{document}

\title{Quasicrystals: Projections of 5-d Lattice into 2 and 3 Dimensions}

\author{Helen AU-YANG and Jacques H.~H.\ PERK}

\address{Department of Physics, \\
Oklahoma State University, \\
Stillwater, OK 74078-3072, USA \\
E-mail: perk@okstate.edu}

\maketitle

\abstracts{
We show that generalized Penrose tilings can be obtained by the projection
of a cut plane of a 5-dimensional lattice into two dimensions, while 3-d
quasiperiodic lattices with overlapping unit cells are its projections into
3d. The frequencies of all possible vertex types in the generalized
Penrose tilings, and the frequencies of all possible types of overlapping
3-d unit cells are also given here. The generalized Penrose tilings are
found to be nonconvertable to kite and dart patterns, nor can they be
described by the overlapping decagons of Gummelt.}

%%%%%%%%%%%%%%%%%%%%%%%%%%%%%%%%%%%%%%%%%%%%%%%%%%%%%%%%%%%%%%%%%%%%%%%%%%%

\section{Introduction}

Quasicrystals, though originally introduced as a mathematical curiosity,
have become an object of intense study by physicists and mathematicians
following the startling discovery in 1984 of five- or ten-fold symmetry in
diffraction patterns off certain alloys.\cite{SBGC} Quasicrystals have been
studied most often by filling the space aperiodically with nonoverlapping
tiles, such as in Penrose tilings.\cite{Penrose,GrSh,Bruijn1} However, in
the mid 1990s, Gummelt\cite{Gummelt} proposed a new description of the
regular Penrose tiling in terms of the overlapping of decorated decagons.
Further research\cite{SJ,SJSTAT,LR,LRK} has shown that this may be a more
sensible way to understand quasicrystalline materials---made of overlapping
unit cells sharing atoms of nearby neighbors.\cite{SJSTAT}

We shall use de Bruijn's multigrid method to produce a new example of
3-dimensional overlapping unit cells.\cite{APoverlap} Moreover, we shall use
the  pentagrid method to obtain generalized Penrose tilings, which cannot be
converted to kite and dart patterns, nor do they satisfy the inflation and
deflation rules. Therefore, since Conway's cartwheels, which are in fact the
overlapping decagons of Gummelt, are constructed from kite and dart
patterns,\cite{GrSh} they cannot be used to describe the generalized Penrose
tilings.

%%%%%%%%%%%%%%%%%%%%%%%%%%%%%%%%%%%%%%%%%%%%%%%%%%%%%%%%%%%%%%%%%%%%%%%%%%%

\section{Grids and the `Cut and Projection Method'}

It is well-known that a Penrose tiling can be obtained by the projection
of a particularly `cut' slice of the 5-d euclidian lattice onto a 2-d
plane ${\cal D}$,\cite{Bruijn1,Bruijn2,GRh} and that its diffraction
pattern,\cite{Elser,DK,Mackay2} therefore, has five- or ten-fold symmetry.
It is also known that not all lattice points $\bk$ in ${\mbox{\mymsbm Z}}^5$
can be mapped onto vertices of a Penrose tiling; only those points in a
particular `cut' slice whose projections into the 3-dimensional orthogonal
space ${\cal W}$ are inside the window of acceptance,\cite{Bruijn2,BJKS}
contribute. The window has been shown\cite {Bruijn2} to be the projection of
the 5-d unit cell Cu(5) with $2^5$ vertices into this 3-d space $\cal W$.

If $\bd_j$ are the generators of the plane ${\cal D}$ and $\bw_j$ are the
generators of its orthogonal space ${\cal W}$, then the projection
operators are the matrices 
\begin{equation}
\bD^\T=(\bd_0,\cdots,\bd_4),\quad \bW^\T=(\bw_0,\cdots,\bw_4)
\label{projoper}
\end{equation}
satisfying $\bD^\T\bW=\bW^\T\bD=0$, where the superscript T denotes matrix
transposition. More specifically, we choose
\begin{equation}
\bd^\T_j=(\cos j\theta,\sin j\theta),\quad
\bw^\T_j=(\cos 2j\theta,\sin 2j\theta,1)=(\bd^\T_{2j},1),
\label{generators}
\end{equation}
where $j=0,\cdots,4$ and $\theta=2\pi/5$. Using notations and ideas
introduced by de Bruijn,\cite{Bruijn1} we consider the 2-d or 3-d
pentagrid consisting of five grids of either equidistant lines given by 
\begin{equation}
x\cos j\theta+y\sin j\theta+\gamma_j=\bd^T_j\br+\gamma_j=k_j,
\quad\br^\T=(x,y), 
\label{grid1}
\end{equation}
or equidistant planes defined by
\begin{equation}
x\cos 2j\theta+y\sin 2j\theta+z+\gamma_j=\bw^\T_j\bR+\gamma_j=k_j,
\quad \bR^\T=(x,y,z),
\label{grid2}
\end{equation}
for $j=0,\cdots,4$, and with the five $k_j\in{\mbox{\mymsbm Z}}$. In
(\ref{grid1}) and (\ref{grid2}),
the $\gamma_j$ are real numbers which shift the grids from the origin. We
denote their sum by
\begin{equation}
\gamma_{0}+\gamma_{1}+\gamma_{2}+\gamma_{3}+\gamma_{4}=c,
\quad 0\le c<1.
\label{shiftc}
\end{equation}
Without loss of generality, we may restrict $c$ to $0\le c<1$, as we can see
from (\ref{grid1}) that $c\to c-n$ if we let $k_0\to k_0+n$. Obviously, such
a relabeling cannot change the 2-d or 3-d quasiperiodic patterns.

It has been shown by de Bruijn\cite{Bruijn1} that the Penrose tiling
associated with a 2-d pentagrid has simple matching rules only for
$c=0$. In other words, for $0<c<1$ the corresponding generalized Penrose
tilings do not satisfy simple matching rules, and have different sets of
vertices for different intervals of $c$.\cite{APwindow} Nevertheless, the
diffraction patterns are believed to be the same for all values of
$c$.\cite{LSt,SoSt}
 
Let the integer $k_j$ be assigned to all points sandwiched between the grid
lines or planes defined by $k_j-1$ and $k_j$. This $k_j$ can be found by
\begin{eqnarray}
&K_j(\br)=\lceil\bd^\T_j\br+\gamma_j\rceil,\quad 
\forall \br\in{\mbox{\mymsbm R}}^2
\label{mesh1}\\
&{\tilde K}_j(\bR)=\lceil\bw^\T_j\bR+\gamma_j\rceil,\quad 
\forall \bR\in{\mbox{\mymsbm R}}^3
\label{mesh2}
\end{eqnarray}
for $j=0,\cdots,4$, (and $\lceil x\rceil$ is the smallest integer greater
than or equal to $x$). A mesh in ${\mbox{\mymsbm R}}^2$ is an interior
area, enclosed by grid lines, containing points with the same five
integers $K_j(\br)$, while a mesh in
${\mbox{\mymsbm R}}^3$ is now an interior volume, enclosed by grid planes,
containing points with the same five integers ${\tilde K}_j(\bR)$. One
then maps each mesh in
${\mbox{\mymsbm R}}^2$ to a vertex in $\cal D$ by 
\begin{equation}
{\myb f}(\br)=\sum_{j=0}^4 K_j(\br)\bd_j=\bD^\T{\myb K}(\br),\quad
{\myb K}^\T(\br)=(K_0(\br),\cdots,K_4(\br)),
\label{map1}\end{equation}
and each mesh in ${\mbox{\mymsbm R}}^3$ to a vertex in $\cal W$ by 
\begin{equation}
{\myb g}(\bR)=\sum_{j=0}^4 {\tilde K}_j(\bR)\bw_j
=\bW^\T{\tilde{\myb K}}(\bR),\quad {\tilde{\myb K}}^\T(\bR)
=({\tilde K}_0(\bR),\cdots,{\tilde K}_4(\bR)).
\label{map2}\end{equation}
The resulting sets of vertices 
${\cal I}=\{{\myb f}(\br)|\br\in{\mbox{\mymsbm R}}^2\}$ and
${\cal L}=\{{\myb g}(\bR)|\bR\in{\mbox{\mymsbm R}}^3\}$ are, respectively,
the two- and three-dimensional quasiperiodic lattices. 

%%%%%%%%%%%%%%%%%%%%%%%%%%%%%%%%%%%%%%%%%%%%%%%%%%%%%%%%%%%%%%%%%%%%%%%%%%%

\section{Window of Acceptance}

Given a point ${\bf k}^T=(k_0,\dots, k_4)$ in the five-dimensional
lattice,\footnote{For a formulation for more general cases, see
Ref.~\refcite{Bruijn2}.} one may ask whether there is a mesh in the
pentagrid (or the 3-d multigrid) such that $K_j(\br)=k_j$ (or
${\tilde K}_j(\bR)=k_j$) for $j=0,\ldots,4$. As seen from (\ref{mesh1})
(or (\ref{mesh2})), this is equivalent to asking whether it is possible to
find points $\br$ in $\mathbb{R}^2$ (or $\bR$ in $\mathbb{R}^3$), and points
${\myb\lambda}^\T=(\lambda_0,\dots,\lambda_4)$ with $0\le\lambda_j<1$,
such that
\begin{eqnarray}
\bD\br+{\myb\gamma}+{\myb\lambda}={\bf k},\label{meshm1},\qquad 
(\bW\bR+{\myb\gamma}+{\myb\lambda}={\bf k}),
\label{meshm}\end{eqnarray}
where ${\myb\gamma}^\T=(\gamma_0,\cdots,\gamma_4)$ and where
${\myb\lambda}$ lies inside the 5-d unit cube Cu(5). Whenever
(\ref{meshm}) holds, the point ${\bf k}$ in ${\mbox{\mymsbm Z}}^5$ is said
to satisfy the mesh condition. Since $\bW^T\bD=\bD^T\bW=0$, the above
equations become
\begin{eqnarray}
&\bW^T[{\bf k}-{\myb\gamma}]=\bW^T{\myb\lambda},
\label{meshw1}\\
&\bD^T[{\bf k}-{\myb\gamma}]=\bD^T{\myb\lambda},
\label{meshw2}
\end{eqnarray}
such that $\bD^T\bk\in{\cal I}$ if (\ref{meshw1}) holds, or
$\bW^T\bk\in{\cal L}$ if (\ref{meshw2}) holds. Thus, $\bW^T{\myb\lambda}$
is the window of acceptance for projections into 2d and
$\bD^T{\myb\lambda}$ for 3d. They are respectively the interiors of the
convex hulls of the points $\bW^\T{\bn_i}$ and $\bD^\T{\bn_i}$, where the
$\bn_i$ are the $2^5$ vertices of the 5-d unit cube Cu(5).

We choose the 32 $\bn_i$'s as follows
\begin{eqnarray}
&\hskip-12pt{\bf n}^T_0\!=\!(0,0,0,0,0),\,
{\bf n}^T_1\!=\!(1,0,0,0,0),\, 
{\bf n}^T_2\!=\!(0,0,0,1,0),\,
{\bf n}^T_3\!=\!(0,1,0,0,0), \hskip-12pt\cr 
&\hskip-12pt{\bf n}^T_4\!=\!(0,0,0,0,1),\,
{\bf n}^T_5\!=\!(0,0,1,0,0),\, 
{\bf n}^T_6\!=\!(1,0,0,1,0),\,
{\bf n}^T_7\!=\!(0,1,0,1,0),\hskip-12pt\cr 
&\hskip-12pt{\bf n}^T_8\!=\!(0,1,0,0,1),\,
{\bf n}^T_9\!=\!(0,0,1,0,1),\, 
{\bf n}^T_{10}\!=\!(1,0,1,0,0),\,
{\bf n}^T_{11}\!=\!(1,1,0,0,0),\hskip-12pt\cr
&\hskip-12pt{\bf n}^T_{12}\!=\!(0,0,0,1,1),\,
{\bf n}^T_{13}\!=\!(0,1,1,0,0), \,
{\bf n}^T_{14}\!=\!(1,0,0,0,1),\,
{\bf n}^T_{15}\!=\!(0,0,1,1,0),\hskip-12pt\cr 
&\hskip-12pt{\bf n}^T_{16}\!=\!(1,1,0,0,1),\,
{\bf n}^T_{17}\!=\!(0,0,1,1,1),\,
{\bf n}^T_{18}\!=\!(1,1,1,0,0),\,
{\bf n}^T_{19}\!=\!(1,0,0,1,1),\hskip-12pt\cr 
&\hskip-12pt{\bf n}^T_{20}\!=\!(0,1,1,1,0),\,
{\bf n}^T_{21}\!=\!(1,1,0,1,0),\,
{\bf n}^T_{22}\!=\!(0,1,0,1,1),\,
{\bf n}^T_{23}\!=\!(0,1,1,0,1),\hskip-12pt\cr
&\hskip-12pt{\bf n}^T_{24}\!=\!(1,0,1,0,1),\,
{\bf n}^T_{25}\!=\!(1,0,1,1,0),\,
{\bf n}^T_{26}\!=\!(1,1,0,1,1),\,
{\bf n}^T_{27}\!=\!(0,1,1,1,1),\hskip-12pt\cr
&\hskip-12pt{\bf n}^T_{28}\!=\!(1,1,1,0,1),\,
{\bf n}^T_{29}\!=\!(1,0,1,1,1),\,
{\bf n}^T_{30}\!=\!(1,1,1,1,0),\,
{\bf n}^T_{31}\!=\!(1,1,1,1,1).\hskip-12pt\cr &\label{vcube}
\end{eqnarray}

The projection of these 32 points into $\cal W$ is a polytope $\cal P$
having 20 faces and 40 edges connecting the 22 vertices, as is shown in
Fig.~1. We let $\bP_i=\bW^\T\bn_i$ for $i=0,\cdots,31$. The bottom is
$\bP_0=(0,0,0)$ and top is $\bP_{31}=(0,0,5)$; they are called the tips of
the polytope. The remaining twenty vertices of $\cal P$ are 
\begin{eqnarray}
&\bP_{j+1}=(\bd_j,1),\quad \bP_{j+6}=(\bd_j+\bd_{j+1},2),\nonumber\\
&\bP_{j+21}=(-\bd_{j-2}-\bd_{j-1},3),\quad \bP_{j+26}=(-\bd_{j-1},4),
\label{hpoly}
\end{eqnarray}
for $j=0,\cdots,4$. The other 10 points $\bP_{11},\cdots,\bP_{20}$
are in the interior of the polytope and are given by
\begin{equation}
\bP_{11+j}=(\bd_j+\bd_{j+2},2),\quad \bP_{16+j}=(-\bd_{j+1}-\bd_{j-1},3),
\label{hpolyi}\end{equation}
again for $j=0,\cdots,4$.

%--------------------------------------------------------------------------
\begin{figure}[ht]
\centerline{\epsfxsize=0.30\hsize\epsfbox{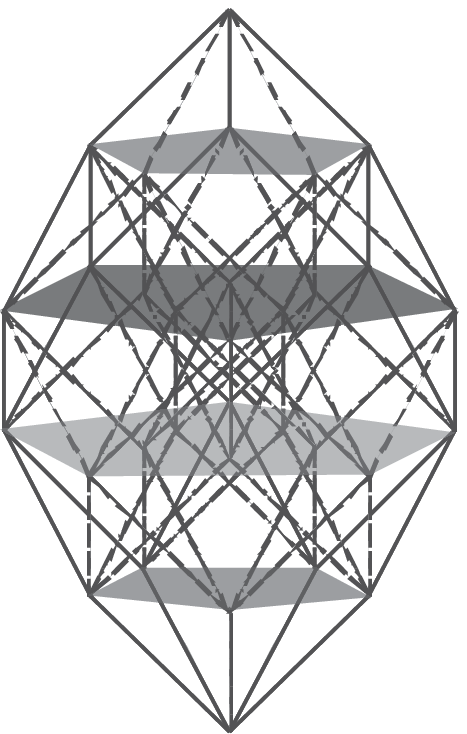}
\hskip50pt
\epsfxsize=0.35\hsize\epsfbox{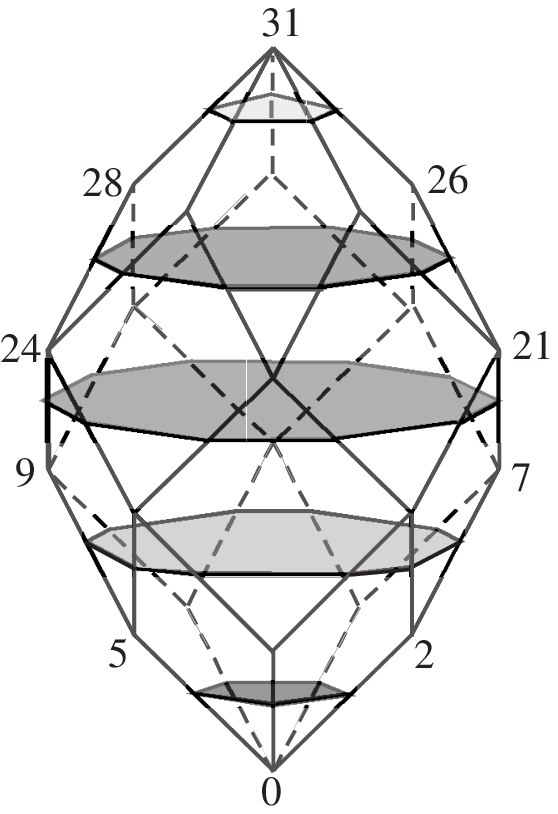}}
\vskip0pt
\hbox to\hsize{\hspace*{60pt}\footnotesize
(a) $c=0$\hspace*{30pt}\hfil\hspace*{40pt}
(b) $c\ne0$\hfil
\hspace*{8pt}}
\vskip0.in\caption{The projection of the 5-dimensional unit cube into the
orthogonal 3-space $\cal W$. The polytopes with 22 vertices are tilted 10
degree with respect to the vertical, so that the intersections $V_I$ with
the planes $z=I-c$ can be seen. In (a), for $c=0$, we show the projection of
the 32 points, 10 of which are in the interior, and the $V_I$ are all
pentagons. In (b), for $c\ne0$, the $V_I$ are pentagons for $I=1,5$, and
decagons for $I=2,3,4$.}
\end{figure}
%--------------------------------------------------------------------------

The orthogonal projection of the 32 points $\bn_i$ into $\cal D$ is a
decagon $\cal Q$ with 10 edges connecting the 10 vertices. Let
$\bQ_i=\bD^\T\bn_i$, for $i=0,\cdots,31$. Then the vertices of the
decagon are
\begin{equation}
\bQ_{11+j}=-p\bd_{3-2j},\quad \bQ_{16+j}=p\bd_{5-2j},
\label{decagon}
\end{equation}
with $j=0,\cdots,4$, and $p=(\sqrt5+1)/2$ is the golden ratio. The
remaining 22 points $\bQ_{0},\cdots,\bQ_{10}$, $\bQ_{21},\cdots,\bQ_{31}$
are in the interior; they are given by
\begin{eqnarray}
&\bQ_{0}=\bQ_{31}=0,\quad \bQ_{j+1}=\bd_{5-2j},\quad
\bQ_{26+j}=-\bd_{2-2j},\nonumber\\
&\bQ_{j+6}=p^{-1}\bd_{4-2j},\quad \bQ_{21+j}=-p^{-1}\bd_{3-2j}.
\label{decagoni}
\end{eqnarray}
The decagons are shown in Fig.~2.
Thus if orthogonal projection $D^\T(\bk-{\myb\gamma})$ is in $\cal Q$,
then its projection $\bW^\T\bk$ is in ${\cal L}$.
%--------------------------------------------------------------------------
\begin{figure}[ht]
\centerline{\epsfxsize=0.6\hsize\epsfbox{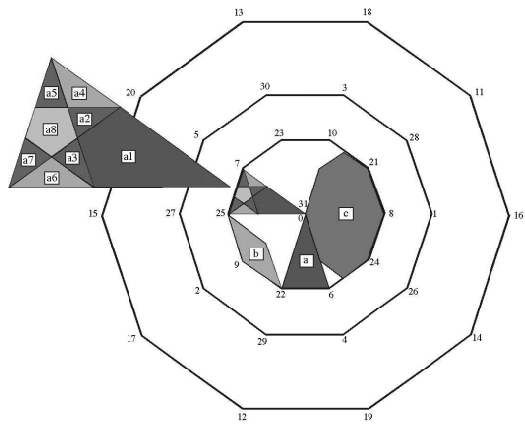}}
\caption{The projection of the 5-d unit cube Cu(5) into the orthogonal 2-d
space $\bD$. The window is a decagon $\cal Q$ whose vertices are given by
(\ref{decagon}). Those $n_i$ which are mapped to interior points (vertices)
of $\cal P$ in Fig.~1, are mapped into the boundary vertices of $\cal Q$.}
\end{figure}
%--------------------------------------------------------------------------
%%%%%%%%%%%%%%%%%%%%%%%%%%%%%%%%%%%%%%%%%%%%%%%%%%%%%%%%%%%%%%%%%%%%%%%%%%%

\section{Generalized Penrose Tilings}

Using (\ref{projoper}) and (\ref{generators}), we may rewrite the three
components of (\ref{meshw1}) as
\begin{equation}
\sum_{j=0}^4(k_j-\gamma_j)=I-c=\sum_{j=0}^4\lambda_j,\quad
\sum_{j=0}^4(k_j-\gamma_j)\bd_{2j}=\sum_{j=0}^4\lambda_j\bd_{2j},
\label{meshc}\end{equation}
where $I\equiv\sum k_j$ is the index of ${\bf k}$, an integer in the
interval $[1,5]$ for $0<c<1$. ($I=5$ does not occur for $c=0$.)
Eq.~(\ref{meshc}) defines the window $V_I$ for accepting ${\bf k}$ with
index $I$. This window $V_I$ is the intersection of the polytope $\cal P$
with the plane at the height $I-c$ shown in Fig.~1.

For $\bk$ in window $V_I$, we examine the condition for its neighbor $\bk'$,
(with $\bk'=\bk\pm\bn_{j}$, $j=1,\cdots,5$), to be in window
$V_{I\pm 1}$. Whenever this condition is satisfied, then $\bD^T\bk$ and
$\bD^T\bk'$ are both vertices of the generalized Penrose tiling. Furthermore
there is a `positive' (`negative') edge incident from the image of $\bk$ in
the direction of $\bd_{3j}$ ($-\bd_{3j}$) to the image of $\bk'$. This way
we can determine all the vertex types of the generalized Penrose tiling
for a given $c$. Denoting all vertices with index $I$ having $n$ `positive'
edges and $n'$ `negative' edges by $[n,n']_I$, we find that for
$\bk\in V_1$ there are only three kinds of vertex types $[5,0]_1$,
$[4,0]_1$, and $[3,0]_1$, and for $\bk\in V_5$ there are also only three
kinds of vertex types $[0,5]_5$, $[0,4]_5$, and $[0,3]_5$, shown in Fig.~3a. 

%--------------------------------------------------------------------------
\begin{figure}[ht]
\centerline{\epsfxsize=0.35\hsize\epsfbox{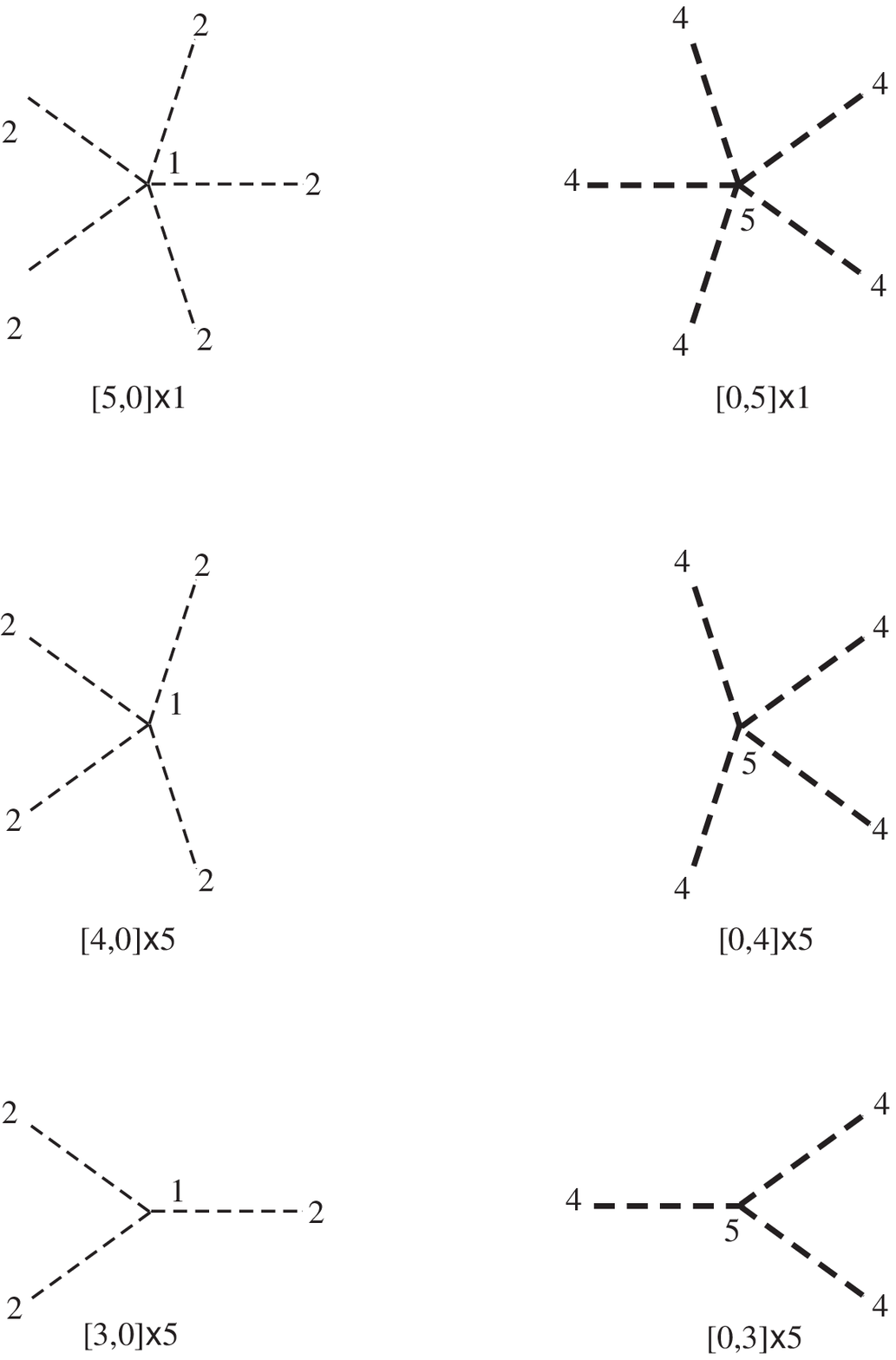}\hskip40pt
\epsfxsize=0.35\hsize\epsfbox{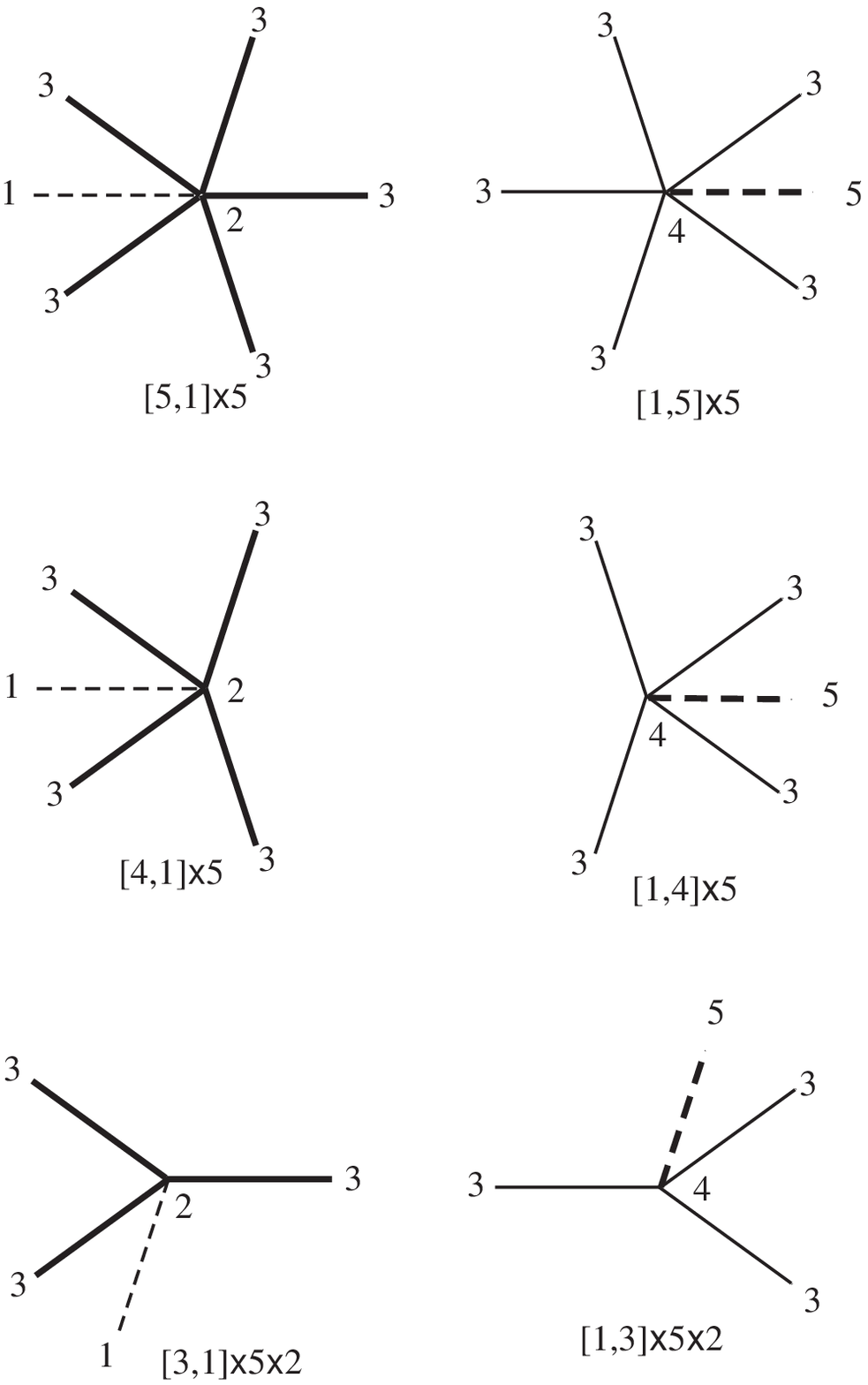}}
\hbox to\hsize{\hspace*{20pt}\footnotesize
(a) vertex types $[n,0]_1$ and $[0,n']_5$\hspace*{40pt}\hfil
(b) vertex types $[n,1]_2$ and $[1,n']_4$\hfil
\hspace*{8pt}}
\vskip0.0in\caption{(a) Edges connecting two sites with indices 1 and 2
are represented by thin dashed lines, while edges connecting sites with
indices 5 and 4 are represented by thick dashed lines. (b) A few examples
of vertex types $[n,1]_2$ and $[n,1]_4$ are given here. Edges
connecting sites with indices 2 and 3 are denoted by thick lines, and
edges connecting sites with indices 3 and 4 by thin lines. We use
$[n,n']\times 5$ to indicate the 5-fold multiplicity under $72^{\circ}$
rotations allowed for the vertex, and $[n,n']\times 5\times 2$ to indicate
the additional reflection symmetry when it is present. }
\end{figure}
%--------------------------------------------------------------------------
If the probability of finding a vertex of type $[n,n']_I$ is denoted by
$A_I(n,n')/5p$, then
\begin{eqnarray}
&A_1[5,0]=\half(p^{-1}\!\!+p)p^{-3}(1-c)^2,\cr
&A_1[4,0]={\textstyle{\frac 5 2}}p^{-4}(1-c)^{2},\,
A_1[3,0]={\textstyle{\frac 5 2}}p^{-3}(1-c)^{2},
\label{Av1}\end{eqnarray}
while $A_5[0,n]$ is given by replacing $1-c$ in $A_1[n,0]$ by $c$.

There are nine different vertex types for $I=2,4$, see Fig.~3b for
some examples of each type. Their frequencies are
\begin{eqnarray}
&A_2[5,0]=\half(p^{-1}\!\!+p)[\theta(p^{-2}\!\!-c)(p^{-3}\!\!+c)^2+
\theta(c-p^{-2})p^{-4}(2-c)^2],\cr
&A_2[5,1]=\theta(p^{-2}\!\!-c){\textstyle{\frac 5 2}}(p^{-5}+c)
p(p^{-2}\!\!-c),\cr
&A_2[5,2]=\theta(p^{-2}\!\!-c){\textstyle{\frac 5 2}}p^{-1}
(p^{-2}\!\!-c)^2,\cr
&A_2[4,0]=\theta(c-p^{-2}){\textstyle{\frac 5 2}}(c-p^{-2})
[p^{-1}(1-c)+p^{-3}(2-c)],\cr
&A_2[4,1]=\theta(p^{-2}\!\!-c){\textstyle{\frac 5 2}}p^{-1}c^2
+\theta(c-p^{-2}){\textstyle{\frac 5 2}}p^{-3}(1-c)^2,\cr
&A_2[3,2]=\theta(p^{-2}\!\!-c){\textstyle{\frac 5 2}}p^2(p^{-2}\!\!-c)^2,\cr
&A_2[3,1]=5p^{-2}(1-c)^2-\theta(p^{-2}\!\!-c)5p^2(p^{-2}\!\!-c)^2,\cr
&A_2[3,0]={\textstyle{\frac 5 2}}c^2-\theta(c-p^{-2})5(c-p^{-2})^2,\cr
&A_2[2,1]={\textstyle{\frac 5 2}}p^{-1}(1-c)^2,
\label{Av2}\end{eqnarray}
where $\theta(x)$ is the Heaviside function, {\it i.e.}, $\theta(x)=1$ for
$x\ge0$, and zero otherwise.
We find that the open interval $0<c<1$ is split into two intervals
$0<c<p^{-2}$ and $p^{-2}<c<1$. Inside the former interval, $A_2(4,0)=0$,
and only eight kinds of vertices are allowed; inside the latter,
$A_2(5,2)=A_2(5,1)=A_2(3,1)=0$, allowing only six vertex types. At the
boundary $c=0$ or $c=p^{-2}$, there are only five allowed vertex types. We
find that $A_4[n,n']$ can be obtained from $A_2[n',n]$ by $c\to 1-c$. Now
for $c$ in the interval $0<c<p^{-1}$ there are six nonvanishing vertex
types, while inside the interval $p^{-1}<c<1$, there are eight nonvanishing
vertex types.

There are many vertex types $[n,n']_3$. Twelve out of twenty of their
frequency functions $A_3[n,n']$ are given as
\begin{eqnarray}
&A_3[0,5]=\theta(p^{-3}\!\!-c)\half(p^{-1}\!\!+p)(p^{-3}\!\!-c)^2,\cr
&A_3[1,5]=\theta(p^{-3}\!\!-c){\textstyle{\frac 5 2}}(p^{-3}\!\!-c)^2,\cr
&A_3[2,5]=\theta(p^{-2}\!\!-c){\textstyle{\frac 5 2}}p^2(p^{-2}\!\!-c)^2
 -\theta(p^{-3}\!\!-c)5p^2(p^{-3}\!\!-c)^2,\cr
&A_3[3,5]=\theta(2p^{-3}\!\!-c)
[{\textstyle{\frac 5 2}}c^2-\theta(c-p^{-3})5p^2(c-p^{-3})^2\cr
&\quad+\theta(c-p^{-2})5p^3(c-p^{-2})^2],\cr
&A_3[4,5]=\theta(c-\!p^{-3})
[\theta(p^{-2}\!\!+p^{-4}\!\!-c)
{\textstyle{\frac 5 2}}p^3(p^{-2}\!\!+p^{-4}\!\!-c)^2\cr
&\quad-\theta(2p^{-3}\!\!-c)5p^3(2p^{-3}\!\!-c)^2
+\theta(p^{-2}\!\!-c)5p^2(p^{-2}\!\!-c)^2],\cr
&A_3[5,5]=\theta(c-p^{-3})
[\theta(2p^{-2}\!\!-c)\half(p+p^{-1})(2p^{-2}\!\!-c)^2\cr
&\quad-\theta(p^{-2}\!\!+p^{-4}\!\!-c)
{\textstyle{\frac 5 2}}p^3(p^{-2}\!\!+p^{-4}\!\!-c)^2
\!\!+\theta(2p^{-3}\!\!-c){\textstyle{\frac 5 2}}p^3(2p^{-3}\!\!-c)^2],\cr
&A_3[3,4]=\theta(p^{-1}\!\!-c)
[{\textstyle{\frac 5 2}}p^{-3}c^2\cr
&\quad-\theta(c-p^{-2})5p(c-p^{-2})^2
+\theta(c-2p^{-3}){\textstyle{\frac 5 2}}p^3(c-2p^{-3})^2],\cr
&A_3[4,4]=\theta(p^{-1}\!\!-c)
[\theta(c\!-\!p^{-2})5(c\!-\!p^{-2})^2
-\theta(c\!-\!2p^{-3})5p^3(c\!-\!2p^{-3})^2\cr
&\quad+\theta(c-\!p^{-2}\!\!-p^{-4})5p^3(c-\!p^{-2}\!\!-p^{-4})^2],\cr
&A_3[3,3]\!=\!5p^{-4}(1\!-\!c)^2\!-\theta(p^{-1}\!\!-\!c)5p^{-1}
(p^{-1}\!\!-\!c)^2\!+\!\theta(p^{-2}\!\!-\!c)5p^{-1}(p^{-2}\!\!-\!c)^2,\cr
&A_3[2,3]=5p^{-3}(1-c)^2
-\theta(p^{-2}\!\!-\!c){\textstyle{\frac 5 2}}p^{-1}(p^{-2}\!\!-\!c)^2,\cr
&A_3[2,2]=10p^{-1}c(1-c),\quad
A_3[1,2]={\textstyle{\frac 5 2}}p^{-1}(1-c)^2. 
\label{Aav3}\end{eqnarray}
The remaining eight $A_3[n',n]$ can be obtained from $A_3[n,n']$ by letting
$c\to 1-c$. They are continuous
functions of $c$.

We plot in Fig.~4 generalized Penrose tilings for
$c=p^{-2}=0.3819660098$ and $c=0.5$. We find that the number of vertices
of index 1 increases, and of index 5 decreases, as $c$ increases.

%%%%%%%%%%%%%%%%%%%%%%%%%%%%%%%%%%%%%%%%%%%%%%%%%%%%%%%%%%%%%%%%%%%%%%%%%%%

\section{Overlapping polytope}

Consider now the projection of ${\mbox{\mymsbm Z}}^5$ into the 3-d space
${\cal W}$. It is easy to find the conditions for both $\bk$ and its
neighbors $\bk+\bn_j$, for $j=1\cdots 5$, to satisfy their mesh conditions,
so that they are vertices of quasiperiodic lattice ${\cal L}$.

We find that every point inside the innermost decagon $\hat{\cal Q}$ in
Fig.~2 corresponds to a point in $\cal L$ that is connected with its 10
neighbors, and is in fact a tip of a polytope. This innermost decagon
$\hat{\cal Q}$ is further divided into 10 triangles. Each point inside a
triangle corresponds to a polytope in ${\cal L}$ having exactly four
interior points which are also in ${\cal L}$. Points in the same triangle
correspond to polytopes having the same four interiors points, but for
different triangles the polytopes have different sets of interior points.
Thus each unit cell contains 26 `atoms,' 22 exterior and 4 interior sites. 

Each of the triangles in $\hat{\cal Q}$ is further divided into eight
regions shown in Fig.~2. The points inside the quadrilateral denoted by
(a1) in Fig.~2, correspond to a polytope intersecting with four other
polytopes and sharing with each a polyhedron $\cal J$ with six faces; inside
the two triangles denoted by (a2) and (a3), each point corresponds to a
polytope intersecting with five other polytopes and sharing with one of them
a polyhedron $\cal K$ with twelve faces and with the other four polyhedra
$\cal J$; inside the two other triangles (a4) and (a6), each point
corresponds to a polytope intersecting with four neighboring polytopes
sharing with one of them a polyhedron $\cal K$ and with the other three
polyhedra $\cal J$; inside the two remaining triangles (a5) and (a7), a
polytope intersects with five other polytopes, sharing with two of them a
polyhedron $\cal K$ and with the other three a polyhedron $\cal J$; inside
the pentagon (a8), a polytope intersects with six other polytopes sharing
with two of them a polyhedron $\cal K$ and with the other four a polyhedron
$\cal J$. Their relative frequencies are related to the ratio of their
areas and are $1:p^{-3}:p^{-2}:p^{-3}:\half(p^{-2}+p^{-4})$. These
frequencies are independent of $c$.
%--------------------------------------------------------------------------
\begin{figure}[ht]
\centerline{\epsfxsize=0.40\hsize\epsfbox{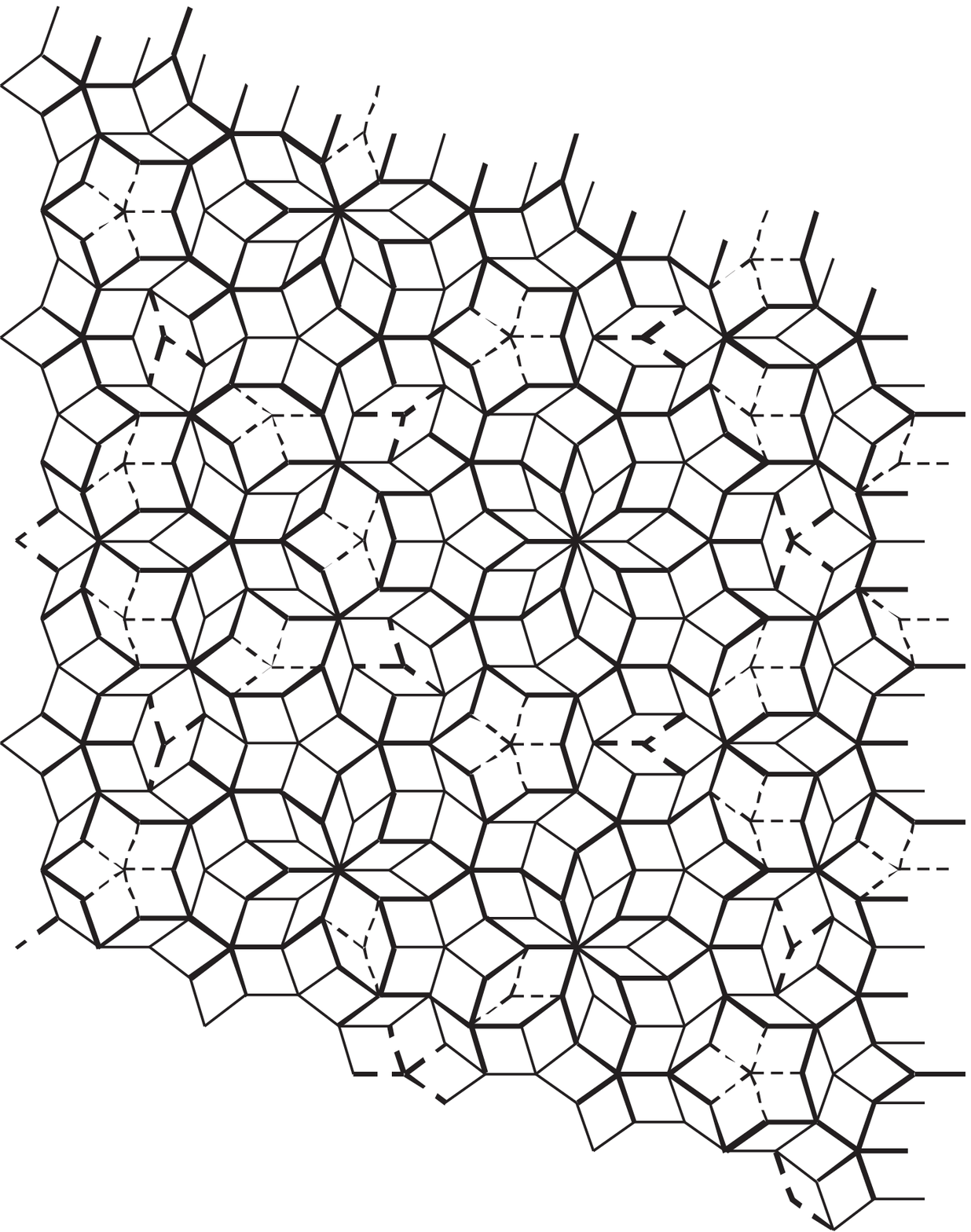}
\hskip40pt
\epsfxsize=0.40\hsize\epsfbox{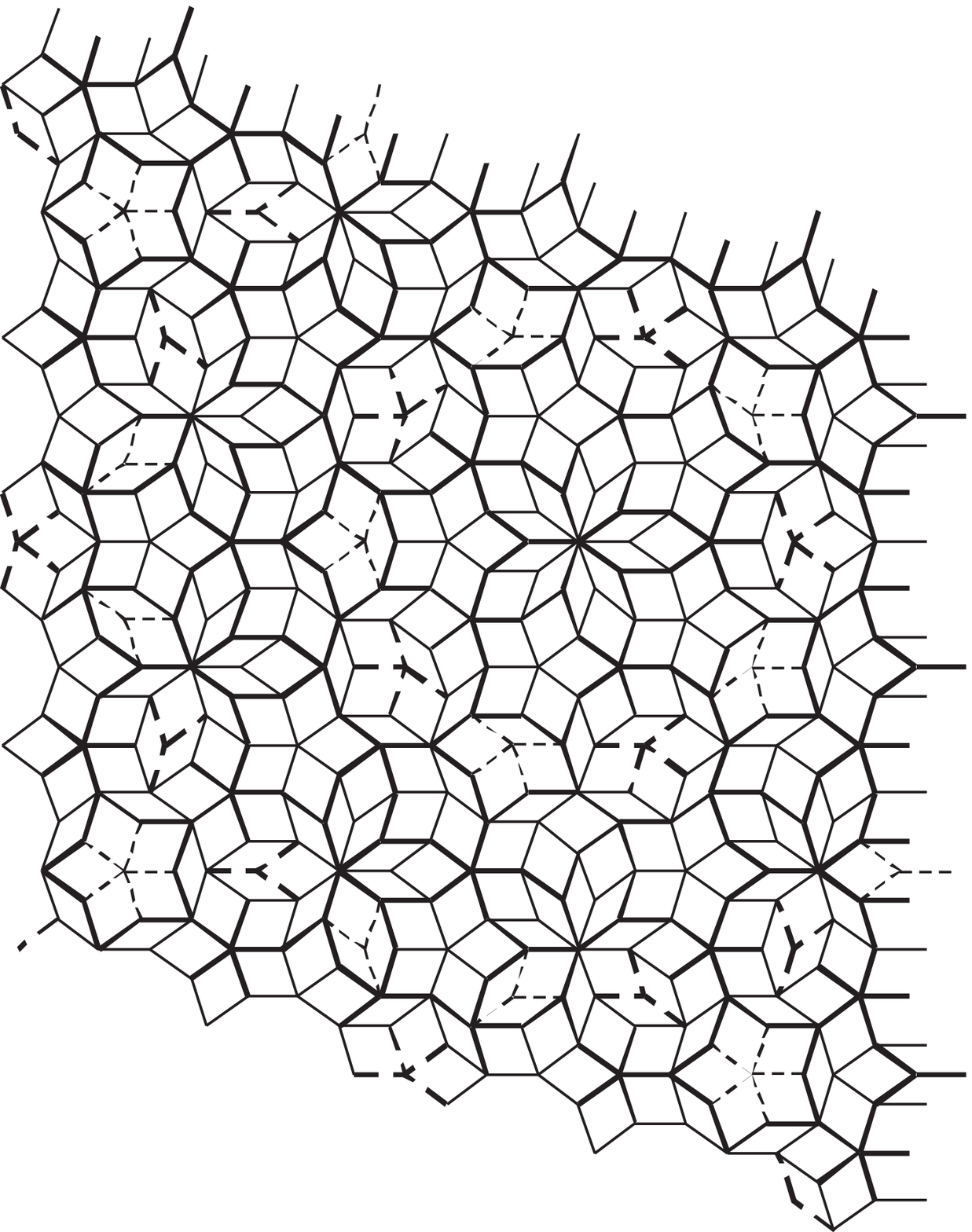}}
\hbox to\hsize{\hspace*{60pt}\footnotesize
(a) $c=p^{-2}$\hspace*{100pt}\hfil
(b) $c=0.5$\hfil
\hspace*{8pt}}
\vskip 0.in
\caption{Generalized Penrose tilings: There are four kinds of edges. Edges
connecting two sites with index 1 and index 2 are represented by a thin
dashed line; edges connecting sites with index 4 and index 5 by a thick
dashed line; edges connecting sites with index 2 and index 3 by a thick
line; and edges connecting sites with index 3 and index 4 by a thin
line. Even though no arrows are drawn on the edges, the `positive'
(connecting $I$ to $I+1$ sites) or `negative' (connecting $I$ to $I-1$
sites) direction of an edge, is completely determined by the indices of the
sites at the two ends of an edge.}
\end{figure}
%--------------------------------------------------------------------------

The 3-d quasiperiodic lattice $\cal L$ can be further shown to be periodic
in the $z$-direction, which is the direction of the line joining the
bottom and the top of the polytopes $\cal P$, and aperiodic in the
$xy$-directions.\cite{APoverlap}

%%%%%%%%%%%%%%%%%%%%%%%%%%%%%%%%%%%%%%%%%%%%%%%%%%%%%%%%%%%%%%%%%%%%%%%%%%%

\section{Conclusion}

The generalized Penrose tilings of thin and fat rhombs cannot be
converted to tilings of kites and darts. This can be seen as follows:
Four thin rhombs and one fat rhomb is the only way to fit the vertex of
type $[3,1]_2$ in Fig.~3b, which can be easily seen to be nonconvertable to
a tiling of darts and kites. On the other hand, for $c=0$, the kite-and-dart
patterns of the Penrose tiling\cite{Gummelt} can be viewed as single
repeating cartwheels,\cite{GrSh} which overlap with their neighbors. These
cartwheels are the overlapping
quasi-unit-cells of Gummelt,\cite{Gummelt,SJ,SJSTAT,LR,LRK} and are larger
than the decagons which are the projections of the 5-d unit cells onto
2 dimensions.\cite{APwindow} The generalized Penrose tilings are shown to be
inequivalent to kite-and-dart patterns, nor do they satisfy the inflation
and deflation rules. Therefore, the method of Gummelt cannot be used here.
It can be seen from Fig.~4 that in the neighborhood of the star vertices
$[5,0]_3$ or $[0,5]_4$, only parts of decagons which are the projections of
the 5-d unit cells onto 2 dimensions\cite{APwindow} are in $\cal L$. This is
not like the case for $c=0$ or for the preojection of the 5-d lattice onto
3-d space. The difference may be due to the fact that 3-d cut hyperplanes in
5d are larger than 2-d cut planes and therefore contain most of
neighboring unit cells Cu(5). For $c=0$, the cut plane for the Penrose
tiling is special such that each decagon which is a projection of the unit
cell Cu(5) into $\cal D$ can also be viewed as quasi-overlapping unit
cell.

%%%%%%%%%%%%%%%%%%%%%%%%%%%%%%%%%%%%%%%%%%%%%%%%%%%%%%%%%%%%%%%%%%%%%%%%%%%

\section*{Acknowledgments}
We are most thankful to Dr.\ M.\ Widom and Dr.\ M.\ Baake for providing us
with many useful references and to Dr. Molin Ge, Dr. Chengming Bai
and the Nankai Institute of Mathematics for their hospitality and support.

%%%%%%%%%%%%%%%%%%%%%%%%%%%%%%%%%%%%%%%%%%%%%%%%%%%%%%%%%%%%%%%%%%%%%%%%%%%


\begin{thebibliography}{00}

\bibitem{SBGC}
D. Shechtman, I. Blech, D. R. Gratias and J. W. Cahn,
%Metallic Phase with Long-Range Orientational
%Order and No Translational Symmetry,
{\it Phys. Rev. Lett.} {\bf 53}, 1951--1953 (1984).

\bibitem{Penrose}
R. Penrose,
%Tilings and Quasi-Crystals; a Non-Local Growth Problem?,
in {\it Introduction to the Mathematics of Quasicrystals},
Aperiodicity and Order, Vol. 2,
M.~V. Jari\'c, ed.,
(Academic, Boston, 1989), pp. 53--79.

\bibitem{GrSh}
B. Gr\"unbaum and G. C. Shephard,
{\it Tilings and Patterns},
(W.~H. Freeman and Co., New York, 1987), Ch.~10.

\bibitem{Bruijn1}
N. G. de Bruijn,
%Algebraic Theory of Penrose's Non-Periodic
%Tilings of the Plane. I,
{\it Indagationes Mathematicae} {\bf 84}, 38--52, % (1981};
%{\rm---}. II,}
%{ibid.}{-}{84},
53--66 (1981).

\bibitem{Gummelt}
P. Gummelt,
%Penrose Tilings as Coverings of Congruent Decagons,
{\it Geometriae Dedicata} {\bf 62}, 1--17 (1996).

\bibitem{SJ}
P. J. Steinhardt and H. C. Jeong,
%A Simpler Approach to Penrose Tilings with implications for
%quasicrystal formation,
{\it Nature} {\bf 382}, 433--435 (1996).

\bibitem{SJSTAT}
P. J. Steinhardt, H. C. Jeong, K. Saitoheong, M. Tanaka, E. Abe
and A. P. Tsai,
%Experimental Verification of the Quasi-Unit-Cell Model of
%Quasicrystal Structure,
{\it Nature} {\bf 396}, 55--57 (1996).

\bibitem{LR}
P. J. Lord and S. Ranganathan,
%The Gummelt Decagon as a `Quasi Unit Cell',
{\it Acta Crystallogr.} {\bf A57}, 531--539 (2001).

\bibitem{LRK}
P. J. Lord, S. Ranganathan and U. D. Kulkarni,
%Quasicrystals: Tiling versus Clustering,
{\it Philosophical Magazine} {\bf A81}, 2645--2651 (2001).

\bibitem{APoverlap}
H. Au-Yang and J. H. H. Perk,
%{\it Overlapping Unit Cells in 3d Quasicrystal Structure},
{\it J. Phys. A} in press, cond-mat/0507117.

\bibitem{Bruijn2}
N. G. de Bruijn,
%Quasicrystal and their Fourier transform,
{\it Indagationes Mathematicae} {\bf A89}, 123--152 (1986).

\bibitem{GRh}
F. G\"ahler and J. Rhyner,
%Equivalence of the Generalised Grid and Projection Methods
%for the Construction of Quasiperiodic Tilings,
{\it J. Phys.} {\bf A19}, 267--277 (1986).

\bibitem{Elser}
V. Elser,
%The Diffraction Pattern of Projected Structures,
{\it Acta Crystallogr.} {\bf A42}, 36--43 (1986).

\bibitem{DK}
M. Duneau and A. Katz,
%Quasiperiodic Patterns,
{\it Phys. Rev. Lett.} {\bf 54}, 2688--2691 (1985).

\bibitem{Mackay2}
A. L. Mackay,
%Crystallography on the Penrose Pattern,
{\it Physica} {\bf A114}, 609--613 (1982).

\bibitem{BJKS}
M. Baake, D. Joseph, P. Kramer and M. Schlottmann,
%Root Lattices and Quasicrystals,
{\it J. Phys.} {\bf A23}, L1037--L1041 (1990).

\bibitem{APwindow}
H. Au-Yang and J. H. H. Perk,
{\it Generalized Penrose Tilings},
to be published.

\bibitem{LSt}
D. Levine and P. J. Steinhardt,
%Quasicrystals. I. Definition and Structure,
{\it Phys. Rev.} {\bf B34}, 596--616 (1986).

\bibitem{SoSt}
J. E. S. Socolar and P. J. Steinhardt,
%Quasicrystals. II. Unit-Cell Configurations,
{\it Phys. Rev.} {\bf B34}, 617--647 (1986).

\end{thebibliography}
\end{document}